\documentclass[twocolumn,footinbib,aps,prl,floats,floatfix,amsmath,superscriptaddress,amssymb,secnumarabic]{revtex4-2}

\usepackage{float}
\usepackage{amsfonts}
\usepackage{amsmath}
\usepackage{slashed}
\usepackage{mathrsfs}
\usepackage{bbm}
\usepackage[scr=rsfs,cal=boondox]{mathalfa}
\usepackage{amssymb}
\usepackage{graphicx}
\usepackage{makeidx}
\usepackage{cancel}
\usepackage{dsfont}
\usepackage{latexsym}
\usepackage{mathtools}
\usepackage[dvipsnames]{xcolor}
\usepackage{float}
\usepackage{multirow}
\usepackage{xurl,hyperref}
\usepackage{tikz-feynman}
\usepackage{enumitem}
\usepackage{soul}
\hypersetup{colorlinks=true,citecolor=red,linkcolor=NavyBlue,urlcolor=NavyBlue}
\usepackage[utf8]{inputenc}
\usepackage{ragged2e}
\usepackage{natbib}
\usepackage{ulem}
\usetikzlibrary{arrows.meta}

\def\ba{\begin{array}}
		\def\ea{\end{array}}

\newcommand{\neut}[1]{\tilde\chi_{#1}^0}

\def\beq{\begin{eqnarray}}
		\def\eeq{\end{eqnarray}}
\begin{document}

\title{Unveiling the Vanishing Higgsino--Nucleon Scattering in the MSSM at Next-to-Leading Order}
\author{Subhadip Bisal}
\email{subhadipbisal6@gmail.com}
\affiliation{School of Physics, Zhengzhou University, Zhengzhou 450000, China}

\author{Arindam Chatterjee}
\email{arindam.chatterjee@snu.edu.in}
\affiliation{Shiv Nadar University, Gautam Buddha Nagar, Uttar Pradesh, 201314, India}

\author{Debottam Das}
\email{debottam@iopb.res.in}
\affiliation{Institute of Physics, Sachivalaya Marg, Bhubaneswar, 751005, India}

\affiliation{Homi Bhabha National Institute, Training School Complex, Anushakti Nagar, Mumbai 400094, India}

\author{Syed Adil Pasha}
\email{sp855@snu.edu.in}
\affiliation{Institute of Physics, Sachivalaya Marg, Bhubaneswar, 751005, India}

\affiliation{Homi Bhabha National Institute, Training School Complex, Anushakti Nagar, Mumbai 400094, India}

\author{Rahul Puri}
\email{rahul.puri@iopb.res.in}
\affiliation{Institute of Physics, Sachivalaya Marg, Bhubaneswar, 751005, India}

\affiliation{Homi Bhabha National Institute, Training School Complex, Anushakti Nagar, Mumbai 400094, India}

\date{\today}

\begin{abstract}
	Higgsino dark matter (DM) is considered one of the most well-motivated and minimal DM scenarios arising from supersymmetric extensions of the Standard Model. Motivated by the requirement of electroweak naturalness, Higgsinos are expected to be relatively light, with masses close to the weak scale. While a pure Higgsino state typically evades current direct detection limits, next-to-leading (NLO) order radiative corrections may bring it within the sensitivity of upcoming experiments. On the contrary, a more important consequence, observed specifically near the kinematic threshold for the production of two particles, is that the NLO corrections lower the DM--nucleon cross section below the neutrino floor. We explicitly examine the cancellation mechanism responsible for suppressed Higgsino--nucleon scattering and identify regions of MSSM parameter space where spin-independent cross-sections may vanish.
\end{abstract}

\maketitle

{\it Introduction.}---Weak-scale supersymmetry (SUSY), in its minimal realisation, the Minimal Supersymmetric Standard Model (MSSM) is endowed with a few remarkable features:
(i) It provides a natural solution to the hierarchy problem associated with the unnaturally small
light Higgs states,
(ii) it offers a well-motivated candidate for cold dark matter (DM) $\tilde \chi_1^0$, namely the lightest superpartner (LSP)
in $R$-parity--conserving scenarios, and
(iii) it can alleviate the issues of flavour and CP violation.
Among various possibilities, a pure Bino LSP has been ruled out by direct searches at the {\sf LHC} \cite{ATLAS:2014jxt, CMS:2015flg, ATLAS:2024exu}.
In contrast, a nearly pure Higgsino (referred as Higgsino-like or $ \tilde{H}$ in this letter) or Wino (Wino-like or $\tilde{W}$) LSP can reproduce the observed DM relic abundance, $\Omega_{\rm DM} h^2 \simeq 0.12$~\cite{Planck:2015bpv, WMAP:2012nax}, through thermal freeze-out for $\mu \sim 1~\text{TeV}$ and $M_2 \sim 2~\text{TeV}$, respectively.
These scenarios are also consistent with current {\sf LHC} bounds, owing to either the typically compressed SUSY spectrum or the heavier masses of other superpartners, which help evade existing collider limits.
Among these possibilities, the $\tilde{W}$ DM scenario faces strong tension with null results from indirect searches for DM annihilation by the {\sf H.E.S.S.}~\cite{Cohen:2013ama, Fan:2013faa} and {\sf MAGIC}~\cite{MAGIC:2022acl} gamma-ray telescopes. On the other hand, Higgsino LSP scenarios remain theoretically well-motivated and are largely unconstrained by current experiments.
In particular, low $\mu$ values, not far from the electroweak (EW) scale, are associated with ``natural" SUSY, as they reduce the degree of fine-tuning required for EW symmetry breaking~\cite{Tata:2020afe}.
Furthermore, the SUSY flavour and CP problems can be mitigated if the first two generations of squarks and sleptons have masses of $\mathcal O(10~{\rm TeV})$, while the third-generation squarks
remain relatively light \cite{Cohen:1996vb}. This can simultaneously preserve naturalness and satisfy flavor and CP constraints.

However, the theoretical prediction of the thermal relic abundance for $\tilde{H}$ DM often leads to a too-small DM density.
The shortfall of the observed abundance in most of the MSSM can be tackled by including new fields
or assuming
non-thermal production of DM.
Direct detection (DD) experiments struggle to probe $\tilde H$ LSPs due to suppressed interaction cross-sections at LO,
as it is proportional to the product of its Higgsino and gaugino components. Similarly, the spin-dependent (SD) cross-section, mediated via $Z$ boson exchange, scales with the square of the difference between the two Higgsino components, $N_{13}^2 - N_{14}^2$, and both channels are further weakened by the typically large mass of the
Higgsino LSP.
However, one-loop corrections, particularly next-to-leading order (NLO) diagrams involving charginos ($\tilde{\chi}_i^\pm$) and $W^\pm$ bosons, can significantly enhance the SI cross-section, even in the nearly pure Higgsino, as these contributions are not directly suppressed by the gaugino content \cite{Drees:1992rr, Drees:1993bu, Drees:1996pk, Hisano:2004pv, Hisano:2010ct, Hisano:2011cs}. Incorporating full EW corrections, the predicted DD cross-section can reach values within the sensitivity of upcoming experiments, such as {\sf Xenon-Lux-Zeplin-Darwin (XLZD)~\cite{XLZD:2024nsu}}, offering the detection of $\tilde H$ DM
as an exciting near-future possibility~\cite{Bisal:2023fgb,  Bisal:2024ezn}.

\begin{figure}[h!]
	\centering
			\begin{tikzpicture}[scale=0.8]
				\begin{feynman}[every edge/.style={line width=0.7pt}]
					\vertex (a1) at (0,1.4) {\(\tilde{\chi}^0_1\)};
					\vertex (a2) at (0,-1.4) {\(q\)};
					\vertex (a3) at (2.4,1.4) {\(\tilde{\chi}^0_1\)};
					\vertex (a4) at (2.4,-1.4) {\(q\)};
					\vertex (b) at (1.2,0.25);
					\vertex (c) at (1.2,-0.85);

					\diagram*{
					(a1) -- [with arrow=0.33, arrow size=1pt] (b) -- [with arrow=0.67, arrow size=1pt] (a3),
					(a2) -- [fermion, arrow size=1pt] (c) -- [fermion, arrow size=1pt] (a4),
					(b) -- [scalar, edge label=${h,H,Z}$, pos=0.7] (c),
					};

					\node[blob,shape=circle, fill=pink, minimum size=0.5cm] at (b) {};
					\node at (1.2,-1.7) {(a)};
				\end{feynman}

			\end{tikzpicture}\hspace{1cm}
			\begin{tikzpicture}[scale=0.8]
				\begin{feynman}[every edge/.style={line width=0.7pt}]
					\vertex (a1) at (7,1.5) {\(\tilde{\chi}^0_1\)};
					\vertex (a2) at (7,-1.5) {\(q\)};
					\vertex (a3) at (10,1.5) {\(\tilde{\chi}^0_1\)};
					\vertex (a4) at (10,-1.5) {\(q\)};
					\vertex (b) at (8.5,0);
					 \vertex (c) at (4,-0.5);

					\diagram*{
					(a1) -- [with arrow=0.3, arrow size=1pt] (b) -- [with arrow=0.7, arrow size=1pt] (a3),
					(a2) -- [with arrow=0.3, arrow size=1pt] (b) -- [with arrow=0.7, arrow size=1pt] (a4),
					};

					draw the blob manually afterwards (this one will show up)
					\node[blob,shape=rectangle, fill=cyan, minimum size=0.8cm] at (b) {};
					\node at (8.5,-1.7) {(b)};
				\end{feynman}
			\end{tikzpicture}
            \\
			\begin{tikzpicture}[scale=0.8]
				\begin{feynman}[every edge/.style={line width=0.7pt}]
					\vertex (a1) at (0,1.4) {\(\tilde{\chi}^0_1\)};
					\vertex (a2) at (0,-1.4) {\(g\)};
					\vertex (a3) at (2.4,1.4) {\(\tilde{\chi}^0_1\)};
					\vertex (a4) at (2.4,-1.4) {\(g\)};
					\vertex (b) at (1.2,0.5);
					\vertex (c) at (1.2,-0.8);

					\diagram*{
					(a1) -- [with arrow=0.33, arrow size=1pt] (b) -- [with arrow=0.67, arrow size=1pt] (a3),
					(a2) -- [gluon] (c) -- [gluon] (a4),
					(b) -- [scalar, edge label=${h,H}$, pos=0.5] (c),
					};

					\node[blob,shape=circle, fill=pink, minimum size=0.4cm] at (b) {};
					\node[blob,shape=circle, fill=green, minimum size=0.4cm] at (c) {};
					\node at (1.2,-1.7) {(c)};
				\end{feynman}
			\end{tikzpicture}\hspace{1cm}
			\begin{tikzpicture}[scale=0.8]
				\begin{feynman}[every edge/.style={line width=0.7pt}]
					\vertex (a1) at (11,1.5) {\(\tilde{\chi}^0_1\)};
					\vertex (a2) at (11,-1.5) {\(g\)};
					\vertex (a3) at (14,1.5) {\(\tilde{\chi}^0_1\)};
					\vertex (a4) at (14,-1.5) {\(g\)};
					\vertex (b) at (12.5,0);

					\diagram*{
					(a1) -- [with arrow=0.3, arrow size=1pt] (b) -- [with arrow=0.75, arrow size=1pt] (a3),
					(a2) -- [gluon,black] (b) -- [gluon,black] (a4),
					};

					\node[blob,shape=circle, fill=magenta, minimum size=0.9cm] at (b) {};
					\node at (12.5,-1.7) {(d)};
				\end{feynman}
			\end{tikzpicture}
	\caption{\justifying Representative diagrams for $\tilde{\chi}_1^0$--$q$ (a) via ($h,H,Z$ exchange) with one-loop corrected vertex, (b) generic four-fermion, along with $\tilde{\chi}_1^0$--$g$ effective interactions (c) via $\neut1 \neut1h/H$ vertex correction, and (d) other irreducible diagrams.}
	\label{fig:sisdschematic}
\end{figure}
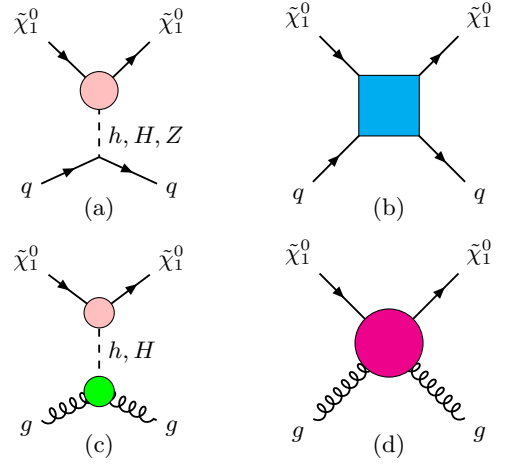

On the contrary, the null results from DD experiments also raise a question about the otherwise well-motivated neutralino and, in particular, for the $\tilde{H}$ DM scenarios.
Thus, it is the purpose of the present article to investigate the criterion to achieve a tiny or even vanishing $\tilde\chi_1^0$--nucleon cross-section, considering the NLO-corrected precision involving MSSM particles.
We schematically represent the contributions in Fig.~\ref{fig:sisdschematic}:
(a) depicts the $\neut1$--$q$ interaction via Higgs exhanges with NLO $\neut1\neut1h_i(Z)$ vertices, (b) shows the box diagram contributions to
the $\tilde \chi_1^0$--$q$
operators, mediated by the $W^\pm$ and $Z$ bosons, (c) shows $\neut1$--$g$ interaction with NLO $\neut1\neut1h_i$ vertices, and (d) illustrates the gluon contributions to the SI interactions, including heavy quark and squark loops at LO (one-loop) and EW particles at NLO (two-loop) processes.
Earlier, the one-loop corrections to the $\tilde{\chi}_1^0$--Higgs vertex mediated by $\tilde \chi_1^\pm$ and $W^\pm$
along with the
quark
and gluon NLO contributions are noted \cite{Hisano:2004pv, Hisano:2010ct,Hisano:2010fy, Hisano:2012wm,
	Hisano:2011cs, Hisano:2015rsa}.
On top of it, in the physical basis, the renormalization of $\tilde{\chi}_1^0$--Higgs
and $\tilde{\chi}_1^0$--$Z$ vertices have been considered to include the contributions from the counterterms for the SI-DD and SD-DD cross-sections
\cite{Bisal:2023fgb, Bisal:2023iip, Bisal:2024ezn,Bisal:2024iar}.
Here, we aim to uncover the cancellation among different
NLO parts, which in turn leads to a tiny or vanishing DM--nucleon scattering cross-section. This tiny cross section is typically associated with so-called \textbf{Blind-Spot} scenarios, in which the DM couplings to the $Z$ or Higgs bosons are suppressed or vanish due to destructive interference among contributing amplitudes~\cite{He:2008qm,He:2011gc,Cheung:2012qy,Chang:2017gla,Huang:2014xua,
	Han:2016qtc,Das:2020ozo}.

\textit{ Effective neutralino--nucleon interactions.}---The relevant
interactions between non-relativistic $\tilde{\chi}_1^0$ and light quarks and gluon at the renormalization scale $\bar{\mu}_0 \simeq m_p$ can be represented as follows \cite{Jungman:1995df, Bertone:2004pz}
\begin{align}
	\mathcal{L}^{\rm eff}
	 & =\eta_q\bar{\tilde{\chi}}^0_1\gamma^\mu\gamma_5\tilde{\chi}^0_1\bar{q}\gamma_\mu\gamma_5 q + \lambda_qm_q\bar{\tilde{\chi}}^0_1\tilde{\chi}^0_1 \bar{q}q \nonumber \\&+\frac{g_q^{(1)}}{m_{\tilde{\chi}^0_1}}\bar{\tilde{\chi}}^0_1 i\partial^\mu\gamma^\nu\tilde{\chi}^0_1\mathcal{O}^q_{\mu\nu}+ \frac{g_q^{(2)}}{m^2_{\tilde{\chi}^0_1}}\bar{\tilde{\chi}}^0_1 (i\partial^\mu)(i\partial^\nu)\tilde{\chi}^0_1\mathcal{O}^q_{\mu\nu}
	\nonumber                                                                                                                                                             \\&+\lambda_G\bar{\tilde{\chi}}^0_1\tilde{\chi}^0_1G^a_{\mu\nu}G^{a\,\mu\nu}~+ ~\dots~.
	\label{eq:Lq_Lg}
\end{align}
The expansion above includes terms up to the second derivative of the neutralino field.
The SD interaction originates from the first term
whereas the SI (coherent) contributions arise from the second and last terms. The third and fourth terms correspond to the quark \textit{twist-2} operators, \textit{i.e.}, the traceless components of the energy-momentum tensor for quarks~\cite{Drees:1993bu, Hisano:2004pv}. These contributions arise from squark-mediated processes at LO and EW box diagrams at NLO. The ellipses denote the analogous gluon twist-2 contributions; both quark and gluon twist-2 contributions are suppressed by heavy squark masses at LO.

Finally, the total DD cross section can be expressed as:
\begin{align}
	\sigma=\frac{4}{\pi}\left(\frac{m_{\tilde{\chi}^0_1}m_T}{m_{\tilde{\chi}^0_1}+m_T}\right)^2
	\Bigg[\{Zf_p+(A-Z)f_n\}^2\nonumber \\+~4\left(\frac{J+1}{J}\right)\{a_p\langle S_p\rangle + a_n\langle S_n\rangle\}^2\Bigg],
	\label{eq:sigma_nucl}
\end{align}
where $m_T$ denotes the mass of the target nucleus, with $Z$ and $A$ representing the atomic and mass numbers of the target nucleus, respectively. Here, $f_N$ ($N=p,n$)
refers to the SI DM--nucleon coupling. In the SD part, the expectation values of the proton and neutron spin contents in the nucleus $\tilde{N}$ at zero momentum transfer are denoted by $\langle S_{N}\rangle = \langle \tilde{N}| S_{N}|\tilde{N} \rangle$. Additionally, $a_N$ denotes the SD DM--nucleon effective coupling.

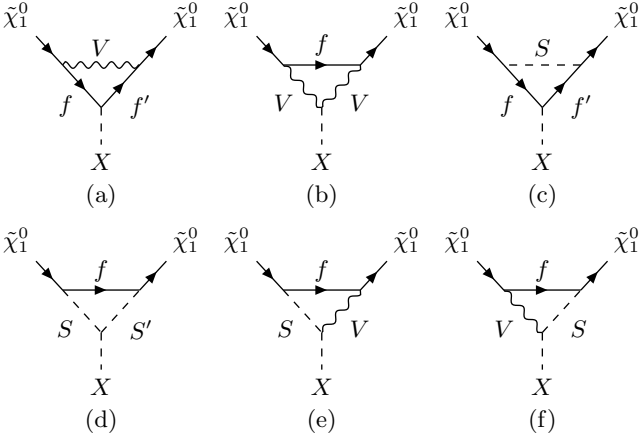
\begin{figure}[h!]
            \newcommand{\scl}{0.51}
            \newcommand{\Xloc}{2.5,-2.1}
            \newcommand{\labloc}{2.5,-3}
            \newcommand{\bbbc}{2.5, -0.7}
            \newcommand{\bbbb}{3.5,0.4}
            \newcommand{\bbba}{1.5, 0.4}
            \newcommand{\lnwdth}{0.5pt}
            
			\begin{tikzpicture}[scale=\scl]
				\begin{feynman}[every edge/.style={line width=\lnwdth}]
					\vertex (a1) at (0.3,1.7) {\(\tilde{\chi}^0_1\)};
					\vertex (b1) at (\bbba);
					\vertex (b2) at (\bbbb);
					\vertex (b3) at (\bbbc);
					\vertex (a3) at (4.7,1.7) {\(\tilde{\chi}^0_1\)};
					\vertex (c) at (\Xloc) {$X$};

					\diagram*{
					(a1) -- [fermion, arrow size=1pt] (b1) -- [fermion, edge label'=$f$, arrow size=1pt] (b3)-- [fermion, edge label'=$f'$, arrow size=1pt] (b2) -- [fermion, arrow size=1pt] (a3),
					(b2) -- [photon, edge label'=$V$] (b1),
					(b3) -- [scalar] (c),
					};
					\node at (\labloc) {(a)};
				\end{feynman}
			\end{tikzpicture}
			\begin{tikzpicture}[scale=\scl]
				\begin{feynman}[every edge/.style={line width=\lnwdth}]
					\vertex (a1) at (0.3,1.7) {\(\tilde{\chi}^0_1\)};
					\vertex (b1) at (\bbba);
					\vertex (b2) at (\bbbb);
					\vertex (b3) at (\bbbc);
					\vertex (a3) at (4.7,1.7) {\(\tilde{\chi}^0_1\)};

					\vertex (c) at (\Xloc) {$X$};

					\diagram*{
					(a1) -- [fermion, arrow size=1pt] (b1) -- [photon, edge label'=$V$] (b3)-- [photon, edge label'=$V$] (b2) -- [fermion, arrow size=1pt] (a3),
					(b1) -- [fermion, edge label=$f$, arrow size=1pt] (b2),
					(b3) -- [scalar] (c),
					};
					\node at (\labloc) {(b)};
				\end{feynman}
			\end{tikzpicture}
			\begin{tikzpicture}[scale=\scl]
				\begin{feynman}[every edge/.style={line width=\lnwdth}]
					\vertex (a1) at (0.3,1.7) {\(\tilde{\chi}^0_1\)};
					\vertex (b1) at (\bbba);
					\vertex (b2) at (\bbbb);
					\vertex (b3) at (\bbbc);
					\vertex (a3) at (4.7,1.7) {\(\tilde{\chi}^0_1\)};
					\vertex (c) at (\Xloc) {$X$};

					\diagram*{
					(a1) -- [fermion, arrow size=1pt] (b1) -- [fermion, edge label'=$f$, arrow size=1pt] (b3)-- [fermion, edge label'=$f'$, arrow size=1pt] (b2) -- [fermion, arrow size=1pt] (a3),
					(b2) -- [scalar, edge label'=$S$] (b1),
					(b3) -- [scalar] (c),
					};
					\node at (\labloc) {(c)};
				\end{feynman}
			\end{tikzpicture}
			\begin{tikzpicture}[scale=\scl]
				\begin{feynman}[every edge/.style={line width=\lnwdth}]
					\vertex (a1) at (0.3,1.7) {\(\tilde{\chi}^0_1\)};
					\vertex (b1) at (\bbba);
					\vertex (b2) at (\bbbb);
					\vertex (b3) at (\bbbc);
					\vertex (a3) at (4.7,1.7) {\(\tilde{\chi}^0_1\)};
					\vertex (c) at (\Xloc) {$X$};

					\diagram*{
					(a1) -- [fermion, arrow size=1pt] (b1) -- [scalar, edge label'=$S$] (b3)-- [scalar, edge label'=$S'$] (b2) -- [fermion, arrow size=1pt] (a3),
					(b1) -- [fermion, edge label=$f$, arrow size=1pt] (b2),
					(b3) -- [scalar] (c),
					};
					\node at (\labloc) {(d)};
				\end{feynman}
			\end{tikzpicture}
			\begin{tikzpicture}[scale=\scl]
				\begin{feynman}[every edge/.style={line width=\lnwdth}]
					\vertex (a1) at (0.3,1.7) {\(\tilde{\chi}^0_1\)};
					\vertex (b1) at (\bbba);
					\vertex (b2) at (\bbbb);
					\vertex (b3) at (\bbbc);
					\vertex (a3) at (4.7,1.7) {\(\tilde{\chi}^0_1\)};
					\vertex (c) at (\Xloc) {$X$};

					\diagram*{
					(a1) -- [fermion, arrow size=1pt] (b1) -- [scalar, edge label'=$S$] (b3)-- [photon, edge label'=$V$] (b2) -- [fermion, arrow size=1pt] (a3),
					(b1) -- [fermion, edge label=$f$, arrow size=1pt] (b2),
					(b3) -- [scalar] (c),
					};
					\node at (\labloc) {(e)};
				\end{feynman}
			\end{tikzpicture}
			\begin{tikzpicture}[scale=\scl]
				\begin{feynman}[every edge/.style={line width=\lnwdth}]
					\vertex (a1) at (0.3,1.7) {\(\tilde{\chi}^0_1\)};
					\vertex (b1) at (\bbba);
					\vertex (b2) at (\bbbb);
					\vertex (b3) at (\bbbc);
					\vertex (a3) at (4.7,1.7) {\(\tilde{\chi}^0_1\)};
					\vertex (c) at (\Xloc) {$X$};

					\diagram*{
					(a1) -- [fermion, arrow size=1pt] (b1) -- [photon, edge label'=$V$] (b3)-- [scalar, edge label'=$S$] (b2) -- [fermion, arrow size=1pt] (a3),
					(b1) -- [fermion, edge label=$f$, arrow size=1pt] (b2),
					(b3) -- [scalar] (c),
					};
					\node at (\labloc) {(f)};
				\end{feynman}
			\end{tikzpicture}
	\caption{\justifying One-loop diagrams for $\neut1\neut1X(h,H,Z)$ vertex correction. Here, $ V\in\{W^\pm,Z\},~f(f')\in\{\neut {1\dots4},\tilde\chi_{1,2}^\pm,\ell,\nu_\ell,q\}$, and $S(S')\in\{h,H,A,H^\pm,G^{0,\pm},\tilde\ell,\tilde\nu_{\ell},\tilde q\}$ represent the MSSM particles.}
	\label{fig:vtx_cxn}

\end{figure}

On top of the LO parts, the total virtual corrections to the $\neut1\neut1h_i(Z)$ from Figs.~\ref{fig:sisdschematic}a and \ref{fig:sisdschematic}c can be written as
\begin{align}
	\Gamma_{\tilde{\chi}_1^0\tilde{\chi}_1^0h_i(Z)} =\sum_{n=a...f}\Gamma_{\tilde{\chi}_1^0\tilde{\chi}_1^0h_i(Z)}^{(n)}
	=C_L^{\rm 1L}\mathbf{P_L} + C_R^{\rm 1L} \mathbf{P_R},
	\label{totalvertex:corrections}
\end{align}
where $\Gamma_{\tilde{\chi}_1^0\tilde{\chi}_1^0h_i(Z)}^{(n)}$ denotes the individual one-loop contributions for the $\tilde{\chi}_1^0\tilde{\chi}_1^0h_i(Z)$ vertex (Fig.~\ref{fig:vtx_cxn}), involving all the MSSM particles. 
For $Z$ mediation, the usual $\gamma^\mu$ vertices are understood and not shown in the diagram.
$C^{\rm 1L}_{L, R}$ is the total one-loop corrections to the coefficients of the left- and right-handed projection operators in the $\tilde{\chi}_1^0\tilde{\chi}_1^0h_i(Z)$ vertex. The complete NLO vertex includes the LO vertex, $C^{\rm LO}_{L, R}$, one-loop virtual corrections $C^{1\rm L}_{L, R}$, and contributions from the counterterms $\delta C_{L, R}$ as
\begin{align}
	C^{\rm NLO}_{L,R}= C^{\rm LO}_{L, R} + C^{\rm 1L}_{L, R} + \delta C_{L, R}.
	\label{eqn:nlo}
\end{align}
The evaluation of counterterms can be achieved through on-shell renormalization, which is employed for the chargino and neutralino sectors in the MSSM~\cite{Eberl:2001eu, Hahn:2015ghv,Chatterjee:2012hkk, Heinemeyer:2011gk, Drees:2006um, Oller:2003ge, Fritzsche:2013fta, Fritzsche:2002bi, Baro:2009gn}. The neutral and charged SUSY fermions are parametrized by the EW gaugino mass parameters $M_1$, $M_2$, and the Higgsino mass parameter $\mu$.
Their mass matrices include the EW gauge boson masses, $\theta_W$, and $\tan\beta$, with all parameters renormalized independently of the chargino and neutralino sectors.
In this way, the mass parameters can be replaced as $M_{1,2}\to M_{1,2}+ \delta M_{1,2}$ and $\mu\to \mu + \delta\mu$.
The counterterms $\delta M_1$, $\delta M_2$, and $\delta \mu$ are determined by imposing on-shell renormalization conditions, such that the masses of $\tilde{\chi}_{1,2}^{\pm}$ and one neutralino $\tilde{\chi}_n^0$ ($n\in\{1, \dots, 4\}$) are given by the poles of their respective propagators. This prescription, known as the {\tt CCN[n]} scheme, where `{\tt C}' and `{\tt N}' signify charginos,  and neutralino respectively, and `{\tt n}' specifies the neutralino $\tilde{\chi}_n^0$ taken on-shell, is used in our analysis.
To ensure numerical stability \cite{Chatterjee:2011wc}, we keep the Bino-like state on shell and therefore adopt the {\tt CCN[3]}  scheme for the $|\mu| < |M_1| < |M_2|$ scenario, as considered here.

We use $\mathtt{FeynArts}$-3.11~\cite{Hahn:2000kx, KUBLBECK1990165, Hahn:2001rv, Fritzsche:2013fta}, $\mathtt{FormCalc}$-9.9~\cite{Hahn:1998yk, Fritzsche:2013fta}, $\mathtt{LoopTools}$-2.16~\cite{Hahn:1998yk}, $\mathtt{SARAH}$-4.14.5~\cite{Staub:2017jnp, Staub:2013tta, Staub:2015kfa}, $\mathtt{SPheno}$-4.0.5~\cite{Porod:2003um, Staub:2017jnp}, and $\mathtt{micrOMEGAs}$-6.2.3~\cite{Belanger:2001fz, Belanger:2006is, Belanger:2008sj, Belanger:2013oya}
at different stages of the computation.

We begin by generating all one-loop and counterterm Feynman diagrams for the $\tilde{\chi}_1^0\tilde{\chi}_1^0h_i(Z)$ vertices and computing the corresponding amplitudes in the Feynman gauge using $\mathtt{FeynArts}$. The loop integrals are then evaluated with $\mathtt{FormCalc}$, and the amplitudes are expressed in terms of Passarino-Veltman (PV) scalar functions.
Next, we compute the renormalisation constants in the {\tt CCN[3]} scheme using $\mathtt{FormCalc}$ and derive the vertex counterterm amplitudes.
We export the analytical expressions for the vertices and counterterms into separate {\tt C++}-functions, then use the MSSM parameter spectrum computed by $\mathtt{SPheno}$ (with $\mathtt{SARAH}$-generated model files) as input for numerical evaluation with $\mathtt{LoopTools}$. The former tools include the effects of the heavy SUSY thresholds to compute physical observables such as Higgs and sparticle masses. After verifying the cancellation of UV divergences, we obtain a finite result for the
$\tilde{\chi}_1^0\tilde{\chi}_1^0 h_i(Z)$ vertex at NLO,
given by Eq.\eqref{eqn:nlo}.
The loop computations employ tree-level masses of the particles to ensure UV finiteness (see, e.g., Refs.~\cite{Bisal:2023fgb, Bisal:2023iip, Bisal:2024ezn,Bisal:2024iar}). We improve the calculation by (a) modifying the $\tilde{\chi}_1^0\tilde{\chi}_1^0h_i(Z)$ vertices in $\mathtt{micrOMEGAs}$
to include all triangular topologies (Figs.~\ref{fig:sisdschematic}a and \ref{fig:sisdschematic}c), (b) including one-loop box diagrams for the $\tilde\chi_1^0$--quark operator (Fig.~\ref{fig:sisdschematic}b), and two-loop contributions to the gluon scalar term $\lambda_G$ (Fig.~\ref{fig:sisdschematic}d). We implement the analytical expressions for the one-loop box and the two-loop scalar gluon contributions from Refs.~\cite{Hisano:2004pv,Hisano:2011cs,Hisano:2012wm}.

\begin{figure}
	\centering
	~~~\includegraphics[width=0.95\linewidth]{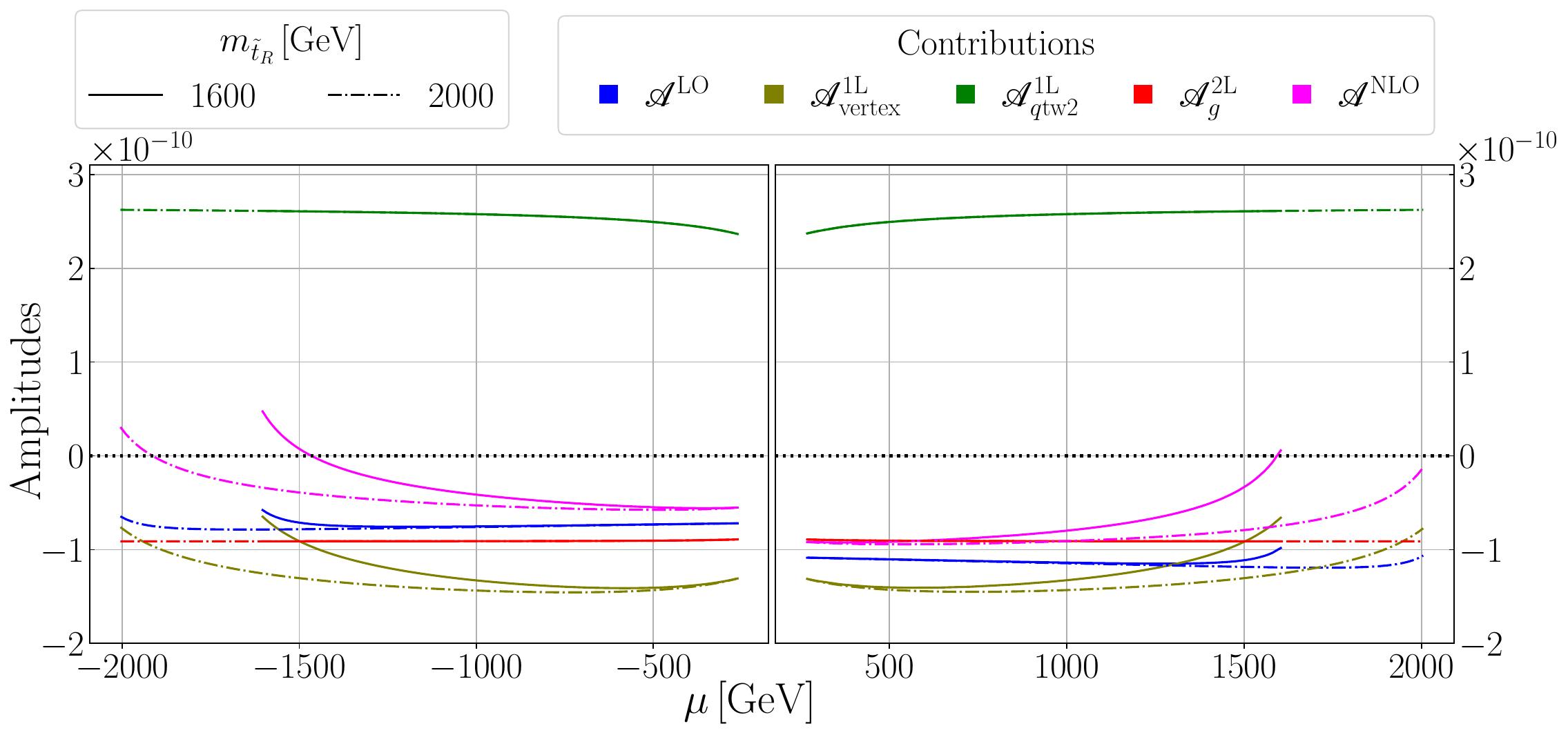}
	\includegraphics[width=\linewidth]{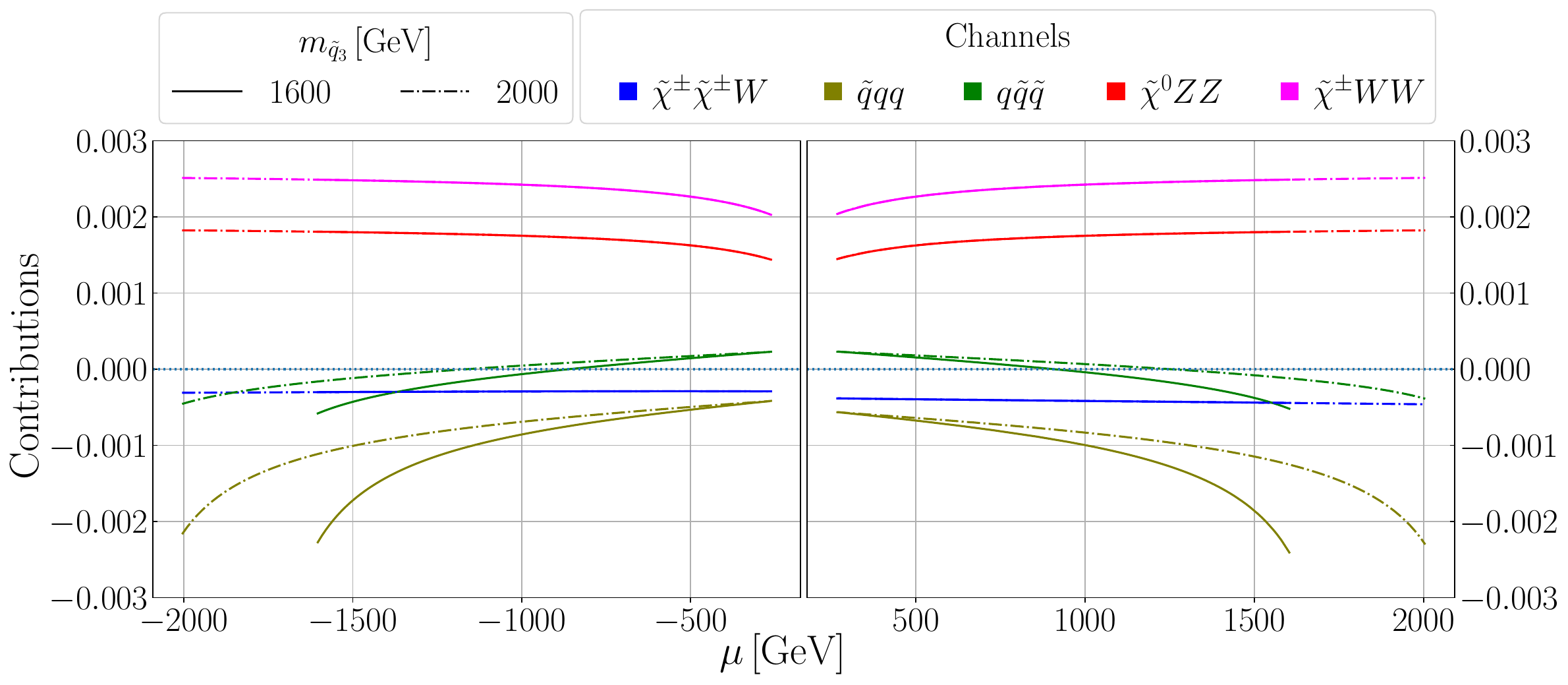}
	\caption{Variation in the contributions to the SI-DD amplitude (top Panel) and to the $\tilde\chi_1^0\tilde\chi_1^0h$ vertex (bottom panel) from various channels, as a function of the $\mu$ parameter for different values of $m_{\tilde t_R}$.}
	\label{fig:strAmp}
\end{figure}

{\it DM--nucleon amplitudes at NLO.}---The LO amplitude receives contributions from tree-level $\neut1\neut1 qq$ and one-loop $\neut1\neut1 gg$ processes, the latter involving both the $hgg$-induced scalar operator and quark–squark loops. The contributions from heavy-Higgs mediated diagrams 
are suppressed in the large-mass limit. For $\tilde H$ DM, the small $\neut1 \neut1 h$ coupling suppresses the scalar contributions in both channels, with a relatively stronger suppression for $\mu < 0$ due to a reduced $\neut1 \neut1 h$ coupling. The $\neut1\neut1h$ vertex-correction contributions, shown in the top panel of Fig.~\ref{fig:strAmp} (olive), correct the $\neut1 \neut1 q q$ amplitude at NLO, with the five largest contributions shown in the bottom panel. The counterterms are not included for illustration. The dominant pieces come from the $\tilde\chi^\pm_1 WW$ (magenta) and $\tilde\chi^0_1 ZZ$ (red) loops, both with the same sign, while the $\tilde qqq$ (olive) and $\tilde q\tilde q q$ (green) loops are subleading.
The NLO contributions to the $\neut{1}\neut{1}gg$ amplitude arise from two-loop diagrams. The dominant terms involve a quark loop combined with a loop containing two $W$ bosons and a chargino ($\tilde{\chi}^\pm_1$).

\begin{figure}[ht!]
	\centering
	\includegraphics[width=\linewidth]{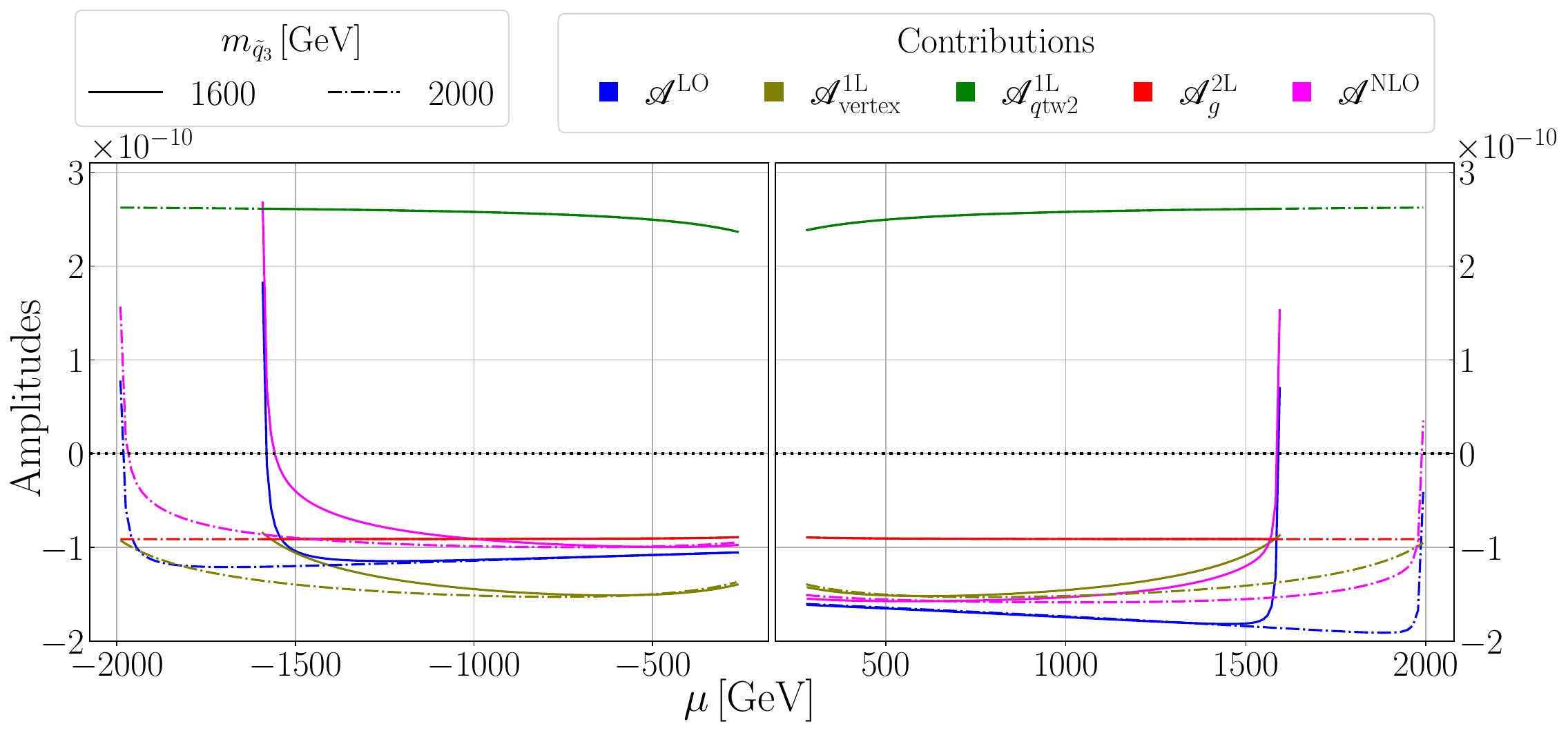}
	\caption{Variation in the contributions to the SIDD amplitude for different values of $m_{\tilde q_3}$.}
	\label{fig:sq3Amp}
\end{figure}

\begin{figure}[h!]
	\centering
	\includegraphics[width=\linewidth]{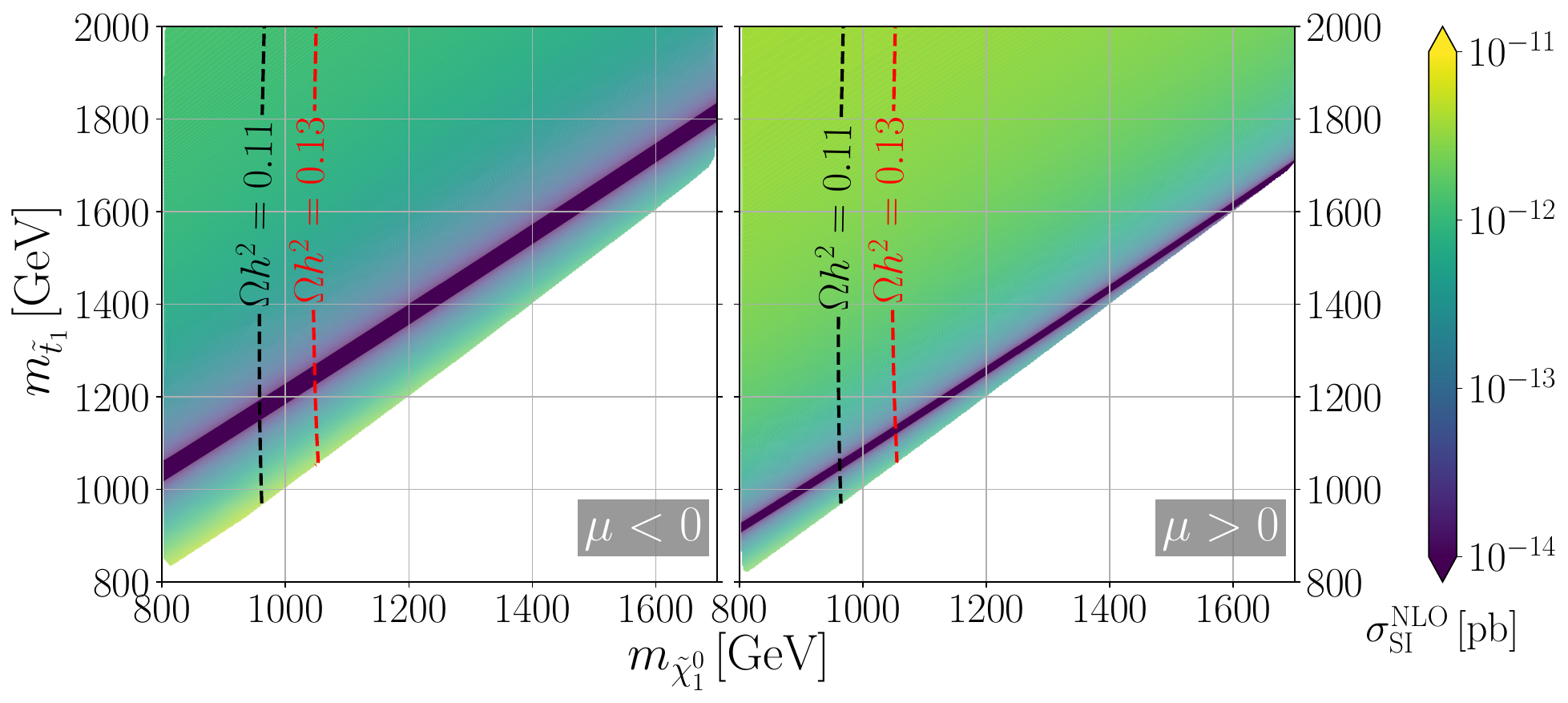}
	\caption{NLO cross-section upon varying $(\mu,m_{\tilde t_R})$ or equivalently ($m_{\neut1},m_{\tilde t_1}$) with $\tan\beta=10$ and  $M_3=M_A=4$\,TeV. Rest of the soft masses are set above 8\,TeV and $T_t=-4$\,TeV.
    }
	\label{fig:2DnegmuBS}
\end{figure}

While the LO contribution ($\mathcal{A}^{\rm LO}$), one-loop corrected $\tilde\chi_1^0\tilde\chi_1^0h$ vertex contribution ($\mathcal{A}^\text{1L}_{\rm vertex}$) and two-loop contributions ($\mathcal{A}^\text{2L}_{g}$) mentioned above interfere constructively, the one-loop quark box contributions ($\mathcal{A}^\text{1L}_{\rm tw2}$) have an overall opposite sign. This destructive interference reduces the total SI amplitude and often leads to \textbf{Blind-Spot} scenarios at NLO~\cite{Bisal:2024iar}. However, for
$\tilde H$ DM, the SI-DD cross-section starts to fall rapidly when
$m_{\neut1}$ approaches the kinematic threshold ($m_{\rm th}=m_{\tilde q}+m_q$) of quark-squark pair 
for a sufficiently light squark. In particular, these threshold can be reached in (I) $\neut1\neut1 h$ three-point vertex corrections through
${\tilde t}$--$t$ threshold (${\tilde b}$--$b$ suppressed via $m_b$) and (II) the $\neut1\neut1 gg$ LO term through
${\tilde q}$--$q$ ($q\in \{u,d,s,c,b,t\}$)
reducing the total amplitude following the destructive interference with $\neut1\neut1 qq$ box contributions. For a light $\tilde t_1$, the former correction dominates the latter due to the large $\tilde t$--$t$ loop contributions.
The top panel of Fig.~\ref{fig:strAmp} shows the dependence of the SI-DD amplitude on $\mu$ for two representative values of $m_{\tilde{t}_R}$, with all other sparticles heavy ($\geq 8$~TeV) and $\tan\beta=10$.
For $\tilde b$--$b$ loop, the threshold neighbourhood can be reached for a lighter LSP, as can be seen in Fig.~\ref{fig:sq3Amp}, where $m_{\tilde q_3}$ is varied.
Similar effects from the loops involving the first two generations of quarks are negligible due to their small Yukawa couplings. 
In general, the MSSM features different kinematic thresholds spanning across all three generations of quark-squark pairs. This in turn can lead to a tiny or even a vanishing DM--nucleon cross-section in different pockets of the MSSM parameter space.
Overall, the sign of $\mu$ produces only mild variations in these corrections, as evident from the Fig.~\ref{fig:strAmp}.
Based on the lesson above, a correlation between the $m_{\neut1}$ and $m_{\tilde q}$ is expected. Fig.~\ref{fig:2DnegmuBS} depicts such a correlation for $\tilde q=\tilde t_1$ where the deep blue region shows $\sigma_{\rm SI}^{\rm NLO}<10^{-14}$~pb.
Here, we take $\tan\beta=10$, $M_3=M_A=-T_t=4$~TeV and rest of the soft masses above 8~TeV.
All of these points satisfy \texttt{SModelS}~\cite{Kraml:2013mwa},
\texttt{HiggsBounds}~\cite{Bechtle:2008jh}, and Higgs mass constraints .

Combining all contributions, we identify the parameter regions where the $\mu$-dependent amplitudes $\mathcal A^{\rm LO}+\mathcal A^{\text{1L}}_{\rm vertex}$ cancel against the nearly $\mu$-independent terms $\mathcal A_{\text{tw2}}^\text{1L}+\mathcal A_{g}^\text{2L}$ for $\tilde H$ DM.

\begin{figure}[h!]
	\centering
	\includegraphics[width=\linewidth]{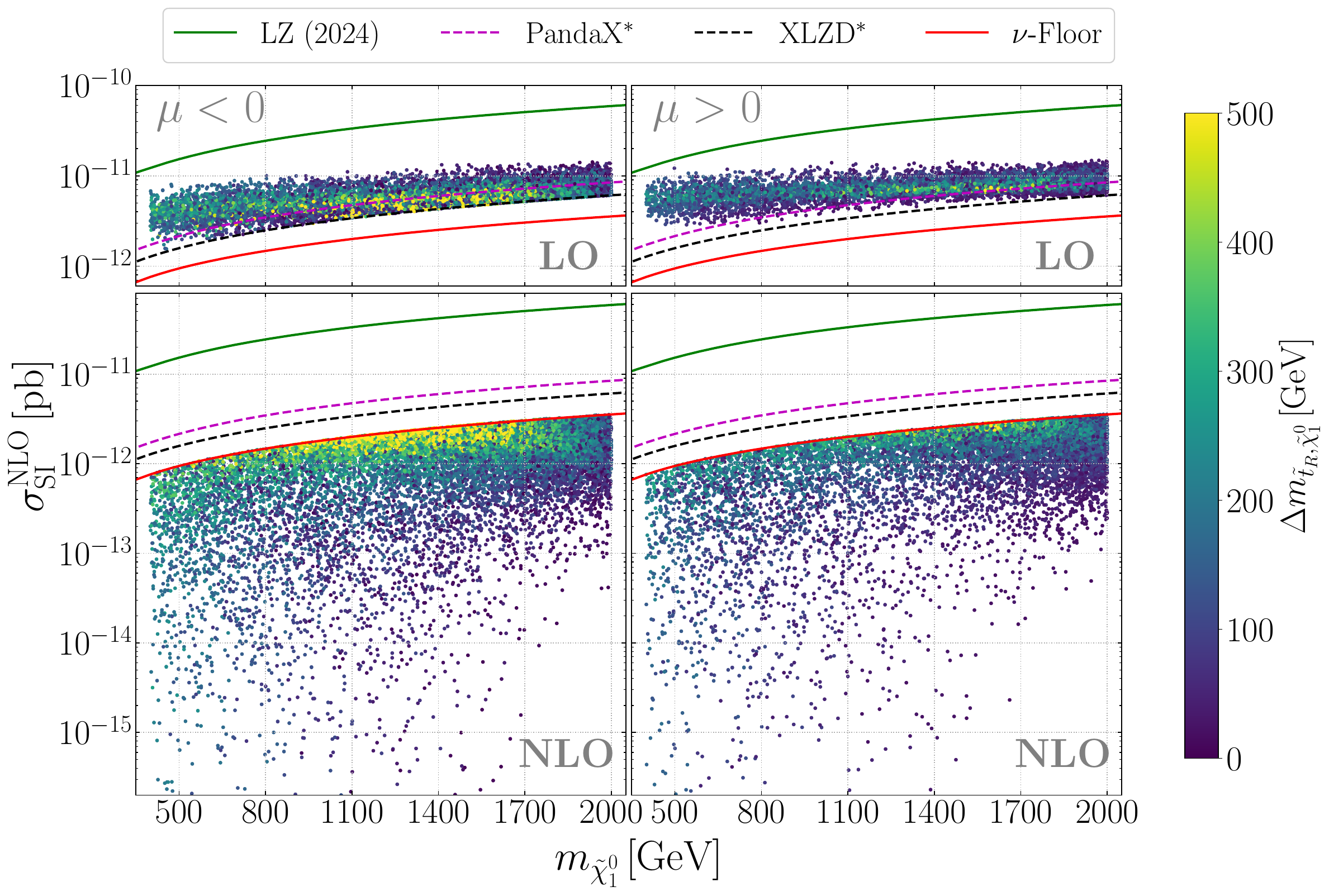}
	\includegraphics[width=\linewidth]{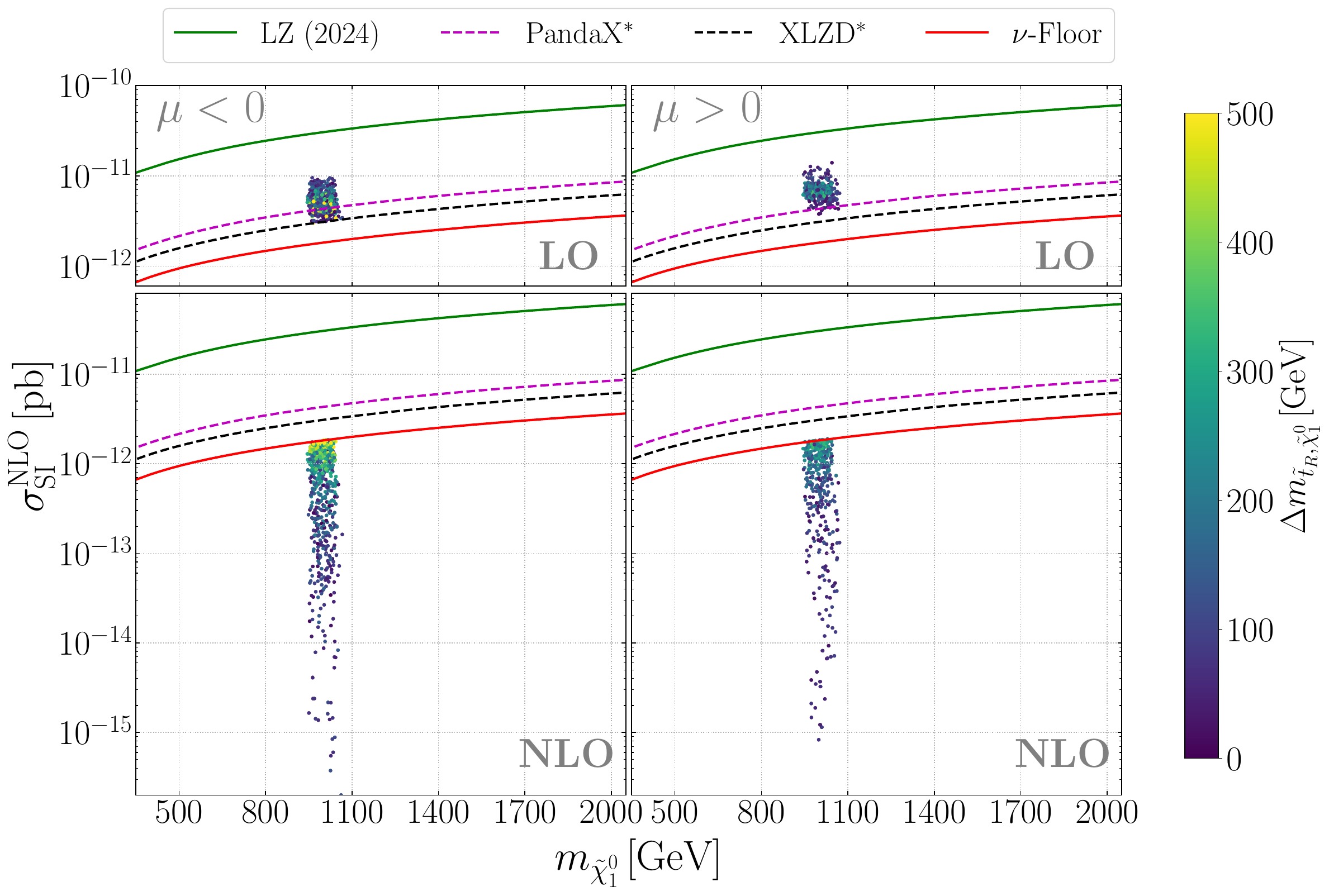}
	\caption{Top Panel: All the points in the scan range which are expected to be ruled out by the limits projected for the {\sf XLZD} collaboration~\cite{XLZD:2024nsu}. The right panel shows points with $\mu>0$, while the left panel contains the points having $\mu<0$. Bottom Panel: The points from the top panel, but with $0.11<\Omega h^2<0.13$. The dashed lines represent the limits projected for the corresponding collaborations. The `$\ast$' marked limits are the projections for upcoming experiments.
    }
	\label{fig:duo}
\end{figure}

For the representation of our results,
we scan the MSSM parameter space for both $\mu>0$ and $\mu<0$ regions by varying the parameters in the ranges:
\begin{align}\small
	~\begin{array}{r@{\,}r@{\,}c@{\,}ll}
		 400\, {\rm GeV} & \leq  & |\mu|             & \leq & 2\, {\rm TeV},\nonumber  \\
		 4 \,{\rm TeV}   & \leq  & M_1               & \leq & 12 \,{\rm TeV},\nonumber \\
		 -4 \,{\rm TeV}  & \leq  & T_t               & \leq & -2 \,{\rm TeV},\nonumber \\
		 1.2\,{\rm TeV}  & \leq  & m_{\tilde{t}_{R}} & \leq & 2.2\,{\rm TeV}.\nonumber \\
		 5               & ~\leq & ~\tan\beta~       & \leq & 50.
	 \end{array}
\end{align}
The other particles are decoupled by taking soft mass parameters high $(\gtrsim10~\rm TeV)$.
In Fig.~\ref{fig:duo}, we present the LO (blue-green-yellow) and NLO (black-red-white) DD cross-sections against $m_{\tilde\chi_1^0}$. Each point shown is excluded at LO by the projected sensitivity of the {\sf XLZD} experiment~\cite{XLZD:2024nsu} but falls below the neutrino floor at NLO, rendering it inaccessible to future DD searches. Additionally, the current limits from the {\sf LUX-ZEPLIN} (2024)~\cite{LZ:2024zvo} experiment, along with the projected sensitivity of the {\sf PandaX} experiment, are also plotted. The color scale represents the difference $\Delta m_{\tilde t_R,\tilde \chi_1^0}\equiv m_{\tilde t_R}-m_{\neut1}$.  All displayed points satisfy $122~{\rm GeV}\le m_{h}\le128~{\rm GeV}$ and the {\tt HiggsBounds} constraints. The left (right) panel corresponds to $\mu<0$ ($\mu>0$), while the lower panel highlights points with loop-corrected masses producing relic density in the range $0.11\le\Omega h^2\le0.13$. 
Our results in Figs.~\ref{fig:2DnegmuBS} and \ref{fig:duo} satisfy the NLO SD cross-section for $\tilde \chi_1^0$--nucleon with $\sigma^{n,p}_{\rm SD}\lesssim10^{-7}$~pb against the current experimental limits of $\sigma^{n}_{\rm SD,LZ}\sim\mathcal O(10^{-6})$~pb and $\sigma^{p}_{\rm SD,LZ}\sim\mathcal O(10^{-4})$~pb for $\sim$1~TeV LSP.

{\it Conclusions.}---
In this letter, we have computed the SI and SD direct-detection cross sections by incorporating all relevant NLO corrections to the parton-level scattering processes $\tilde{\chi}_1^0\tilde{\chi}_1^0\bar{q}q$ and $\tilde{\chi}_1^0\tilde{\chi}_1^0gg$ for $\tilde H$ DM in the context of natural SUSY.
Various contributions to the $\tilde H$--nucleon scattering amplitude, arising from tree-level, one-loop triangular and box diagrams as well as two-loop diagrams, exhibit significant cancellations, which in turn markedly suppress the predicted cross sections. Thus, regions that would appear testable even by XLZD at LO can become effectively unreachable after NLO corrections. For a thermal $\tilde H$ DM,
the cancellations become stronger when heavy quark–squark loops in $\tilde{\chi}_1^0\tilde{\chi}_1^0\bar{q}q$, and $\tilde{\chi}_1^0\tilde{\chi}_1^0 gg$ approach the threshold of two-particle production for a non-relativistic $\tilde\chi_1^0$.
Consequently, the SI cross-section may fall well below $\mathcal{O}(10^{-14})$~pb, deep into the neutrino fog.

Being several orders of magnitude smaller than current direct detection limits, these values are congruent with the absence of a signal. Furthermore, we identify a substantial region of parameter space that is compatible with existing experimental bounds and therefore opens the next window for testing SUSY models, such as the MSSM.

The computations in this project were partially supported by SAMKHYA, the high-performance computing (HPC) facility provided by the Institute of Physics, Bhubaneswar (IOPB). The authors acknowledge S. Vempati for the valuable discussions.

\bibliography{ref}
\end{document}